\newtheorem{definition}{Definition}

\newtheorem{remarque}{Remark}

\documentclass{acm_proc_article-sp}
\usepackage[french,linesnumbered,algoruled]{algorithm2e}
\SetAlFnt{\small\sf}
\begin{document}


%
%
%



\title{Nouvelle approche de recommandation personnalisée dans les folksonomies basée sur le profil des utilisateurs}

\numberofauthors{3}
\author{
\alignauthor
Mohamed Nader Jelassi\\
       \affaddr{Université de Tunis El-Manar, Faculté des Sciences de Tunis, LIPAH,
2092, Tunis, Tunisie}\\
       \affaddr{Clermont Université, Université Blaise Pascal, LIMOS, BP 10448, F-63000
Clermont Ferrand}\\
       \affaddr{nader.jelassi@isima.fr}
\alignauthor Sadok Ben Yahia\\
     \affaddr{Université de Tunis El-Manar, Faculté des Sciences de Tunis, LIPAH,
2092, Tunis, Tunisie}\\
       \affaddr{sadok.benyahia@fst.rnu.tn}
\alignauthor Engelbert Mephu Nguifo\\
      \affaddr{Clermont Université, Université Blaise Pascal, LIMOS, BP 10448, F-63000
Clermont Ferrand}\\
       \affaddr{engelbert.mephu\_nguifo@univ-bpclermont.fr}
}

\maketitle

\begin{abstract}
Depuis des années, la popularité et simplicité des folksonomies a incité de plus en plus d'utilisateurs à partager des ressources en les annotant par des tags de leurs choix. Les utilisateurs qui sont les principaux acteurs du système possèdent des profils différents qui doivent être pris en compte par le système afin de leur proposer des recommandations de tags, de ressources ou d'amis plus personnalisées. Dans ce papier, nous considérons le profil de l'utilisateur comme quatrième dimension d'une folksonomie, classiquement composée de trois dimensions (utilisateurs, tags, ressources) et nous proposons une approche de regroupement des utilisateurs au profil et intérêts équivalents sous forme de concepts quadratiques. Par suite, nous utilisons ces concepts pour proposer notre système personnalisé de recommandation de tags, de ressources ou d'amis selon le profil de l'utilisateur. Les expérimentations menées sur le jeu de données \textsc{MovieLens} a donné des résultats encourageants en termes de précision.
\end{abstract}

\section{Introduction}

Une \emph{folksonomie} est un néologisme, né de la jonction des mots \emph{folk} (\emph{i.e.,} les gens) et \emph{taxonomie}, désignant un système de classification collaborative par les internautes \cite{Mika2005aInf}. L'idée est de permettre à des utilisateurs de partager et de décrire des objets via des tags librement choisis. Formellement, une \emph{folksonomie} est composée de trois ensembles $\mathcal{U}$, $\mathcal{T}$, $\mathcal{R}$ et d'une relation ternaire $\mathrm{Y}$ entre eux, où $\mathcal{T}$ est un ensemble de tags (ou étiquettes) et $\mathcal{R}$ est un ensemble de ressources partagées par les utilisateurs, qui peuvent être des sites web à marquer\footnote{\emph{http://del.icio.us}}, des vidéos personnelles à partager\footnote{\emph{http://youtube.com}} ou des films à décrire\footnote{\emph{http://movielens.org}} selon le type de la \emph{folksonomie} considérée \cite{hotho2006information}. Quant à l'ensemble $\mathcal{U}$, il consiste en l'ensemble d'utilisateurs d'une \emph{folksonomie} qui sont décrits par leurs identifiants (pseudonymes). Les utilisateurs sont les acteurs principaux du système étant donné qu'ils contribuent au contenu par l'ajout de ressources et l'affectation de tags: ils sont donc considérés comme les créateurs de l'information. La participation massive et croissante des utilisateurs dans les \emph{folksonomies} revient au fait que la participation au contenu ne nécessite aucune connaissance spécifique \cite{Jaschke08}, \emph{i.e.,} chacun est capable de contribuer au contenu sans beaucoup d'efforts donnant le plein pouvoir aux utilisateurs, acteurs principaux de la \emph{folksonomie}.

Cependant, il s'avère que le choix de tags et de ressources partagées par un utilisateur d'une \emph{folksonomie} dépendait de son profil: le genre, l'âge, la profession, etc. \cite{Michlmayr}. Cette diversité culturelle résultante des \emph{folksonomies} est tout à fait fascinante. Si cette diversité peut être considérée comme un point fort des \emph{folksonomies}, cela peut également être vu comme un point faible. En effet, les \emph{folksonomies} ont à tenir compte du profil de chacun lors de la recommandation de tags ou de ressources, ce qui n'est pourtant pas le cas. Cette lacune a incité les chercheurs à proposer des systèmes de recommandation personnalisés afin de suggérer les tags et ressources les plus personnalisées et appropriés au profil des utilisateurs. Pour illustrer cela, considérons l'utilisateur \emph{Mike}, un étudiant appartenant à la catégorie d'âge des "$18$-$25$ ans", notre objectif est de lui suggérer des tags et ressources les plus partagés par les utilisateurs au même profil (étudiants de la même catégorie d'âge), \emph{i.e.,} utilisant le même vocabulaire et intéressés par les mêmes ressources dans la \emph{folksonomie}. Par ailleurs, nous pouvons lui proposer des utilisateurs (amis) ayant un profil et des intérêts équivalents, \emph{i.e.,} s'intéressant aux mêmes tags et ressources.

Pour atteindre cet objectif, nous considérons le profil des utilisateurs comme une quatrième dimension d'une \emph{folksonomie}, classiquement composée de trois dimensions (utilisateurs, tags et ressources), et nous proposons une approche de regroupement des utilisateurs au profil et intérêts équivalents sous forme de structure appelées concepts quadratiques\cite{jelassi}. Notre approche est basée sur une extension de l'algorithme \textsc{TriCons} \cite{pakdd}, qui est dédié à l'extraction des concepts triadiques à partir des \emph{folksonomies}, pour l'extraction de concepts quadratiques d'utilisateurs, tags, ressources et profils. De telles structures permettent donc de regrouper à la fois des informations sur l'historique de tagging et sur le profil des utilisateurs. Par exemple, un concept quadratique possible peut être: \emph{"les $18$-$25$ ans utilisent les mêmes tags pour annoter les films d'action et d'aventure"}. Par suite, nous pouvons proposer un système personnalisé de recommandation de tags et de ressources selon chaque profil d'utilisateur. En outre, nous sommes en mesure de suggérer une liste d'utilisateurs (amis) au même profil et aux intérêts équivalents \cite{theseliang}.



Le reste du papier est organisé comme suit: dans la Section \ref{sec2}, nous motivons notre groupement d'utilisateurs au même profil dans le but de suggérer les tags et ressources les plus en concordance avec le profil des utilisateurs. Nous étudions les principales approches de la littérature dans la Section \ref{sec3}. Dans la Section \ref{sec4}, nous introduisons une définition formelle d'une \emph{folksonomie} et d'un quadri-concept. Dans la Section \ref{sec5}, nous proposons un système personnalisé de recommandation. Ensuite, nous menons une étude expérimentale avec pour principal objectif la suggestion d'utilisateurs, tags et ressources dans la Section \ref{sec6}. Enfin, nous concluons notre papier et donnons quelques perspectives pour nos travaux futurs dans la Section \ref{sec7}.

\section{Motivations} \label{sec2}

   L'essor des \emph{folksonomies}, dû principalement au succès des systèmes sociaux de partages de ressources (\emph{e.g.,} \textsc{Flickr}, \textsc{Bibsonomy}, \textsc{Youtube}, etc.), a suscité l'intérêt des chercheurs pour la recommandation. En effet, le manque d'organisation des ressources partagées ainsi que la (trop) grande liberté de choix de tags ont incité les travaux actuels à améliorer les recommandations actuelles afin d'aider l'utilisateur à choisir les "bonnes" ressources (les plus intéressantes parmi les milliers de ressources disponibles) à partager et à utiliser les "bons" tags pour les annoter. Cela a entraîné le besoin de personnaliser ces recommandations. Mais pourquoi a-t-on si besoin de personnaliser? \cite{Michlmayr} \cite{landia} \cite{Vallet} Les utilisateurs d'une \emph{folksonomie} ont des profils et des attentes différentes qui dépendent de leurs motivations. C'est dans un souci de répondre aux besoin de chaque utilisateur que des travaux se sont intéressés à la personnalisation des recommandations. Cependant, pourquoi, dans notre cas, a-t-on besoin de connaître en outre le profil de l'utilisateur? \cite{coria2} \cite{Noll} Pour réussir ou tenter de répondre au mieux aux attentes de chaque utilisateur de la \emph{folksonomie}, il est utile d'avoir plus d'informations sur lui \cite{whotags}. En effet, son âge, sa profession ou sa localisation sont des informations qui sont susceptibles de nous aider dans le processus de personnalisation de recommandation. Prenons par exemple le cas de quatre utilisateurs (un homme riche, un zoologue, une adolescente et un étudiant de $20$ ans) cherchant des ressources correspondant au tag \emph{Jaguar}. Selon le profil de chacun, une recommandation différente (page web, photos, $\ldots$) lui sera proposée. Ainsi, le premier utilisateur aura comme recommandations des ressources sur les voitures de marque \emph{Jaguar}; le zoologue aura des ressources recommandées sur le \emph{Jaguar}, l'animal. De même, il est plus intéressant de recommander des ressources sur le groupe de musique anglais \emph{Jaguar} pour l'adolescente. Enfin, nous pourrons suggérer des ressources sur le film \emph{Le Jaguar} pour l'étudiant de $20$ ans).

   Ceci étant, de quelle manière le recours aux concepts quadratiques va nous aider à parvenir à cet objectif? D'un côté, si on peut facilement étudier les tags utilisés par un seul utilisateur sur une ressource, il est évident de constater que la tâche devient rapidement intraitable pour un ensemble de taggings impliquant plusieurs utilisateurs et plusieurs ressources. D'un autre côté, les tags (ou ressources) recommandés s'avèrent ne pas être très spécifiques \cite{tagrecomm6}, \emph{i.e.,} des tags qui sont des mots "bateau" ou bien des ressources vagues ne correspondant pas au besoin spécifique de l'utilisateur. Grâce aux concepts quadratiques, nous pouvons résoudre ces deux problèmes. En effet, d'un côté, les concepts quadratiques sont des strucutres regroupant les tags et ressources en commun à un ensemble maximal d'utilisateurs ayant les mêmes profils. D'un autre côté, dans un concept quadratique, les tags et ressources qui ont été utilisés en combinaison seront regroupés d'où un résultat plus spécifique. Formellement, un concept quadratique est un quadruplet ($U$, $T$, $R$, $P$) formé d'un ensemble $U$ d'utilisateurs, d'un ensemble $T$ de tags, d'un ensemble $R$ de ressources et d'un ensemble $P$ de profils \cite{datapeelertkdd}. Dans un concept quadratique, les utilisateurs ayant un profil équivalent et qui partagent des tags et des ressources en commun sont alors regroupés ensemble. Une fois extraits, ces quadri-concepts sont utilisés pour notre algorithme de recommandation personnalisée axé sur les trois domaines suivants:

 \begin{itemize}
   \item Un utilisateur, avec un profil donné partageant une ressource donnée, aura une suggestion personnalisée de tags correspondant à son vocabulaire.
   \item La recommandation de ressources correspondant au profil d'un utilisateur cible.
   \item La suggestion d'utilisateurs au profil et intérêts équivalents qu'un utilisateur cible.
 \end{itemize}

\subsection*{Contribution}
Dans ce papier, nous avons étendu la \emph{folksonomie} en considérant le profil des utilisateurs comme une nouvelle dimension. L'extension de la \emph{folksonomie} par les profils utilisateurs est intéressante dans le sens où nous aurons plus d'informations pour le processus de recommandation. Nos recommandations personnalisées sont donc réalisées afin d'éviter d'offrir la même recommandation à tous les utilisateurs indépendamment de leurs profils. La principale originalité de notre approche est le recours aux concepts quadratiques (extraits à partir de la \emph{folksonomie} étendue) qui sont des structures qui regroupent des ensembles maximaux d'utilisateurs, tags et ressources. Ainsi, notre algorithme de recommandation se base à la fois sur le profil des utilisateurs et sur leurs tagging afin de recommander les tags et ressources partagés les plus spécifiques.
Dans ce qui suit, nous menons une étude critique des différentes approches de la littérature qui ont proposé de personnaliser les recommandations.

\section{Travaux connexes} \label{sec3}
Dans un souci d'améliorer les recommandations dans les \emph{folksonomies}, plusieurs travaux ont été proposés dans la littérature.

 Dans \cite{commun}, les auteurs utilisent la \emph{"personomie"} d'un utilisateur, \emph{i.e.,} les tags qui lui sont relatifs, afin de lui recommander des utilisateurs ayant partagé des tags et ressources similaires. Tout d'abord, ils construisent un profil basé sur les tags les plus fréquents pour un utilisateur donné, \emph{i.e.,} ceux qui sont les plus utilisés. Ensuite, à partir de ce profil, les auteurs sont capables de recommander des utilisateurs (dits \emph{collaborateurs}) en utilisant une mesure de similarité entre utilisateurs appelée \emph{cos\_iuf}. Cette approche, qui se focalise sur une seule dimension (les utilisateurs), souffre d'une mesure plus qu'approximative. Cette mesure, qui s'appuie uniquement sur les tags utilisés par les utilisateurs, n'offre pas une information complète sur les utilisateurs. Plus récemment, dans \cite{socialcontacts}, les auteurs se basent à la fois sur l'historique de tagging des utilisateurs et sur leurs préférences personnalisées (à partir de leurs contacts sociaux). Selon les auteurs, les données relatives aux contacts sociaux d'un utilisateur peuvent donner des recommandations de tags plus personnalisées lors de l'opération de tagging. La limite de cette approche est qu'elle requiert qu'un utilisateur doit posséder des contacts sociaux afin d'avoir des recommandations de tags. Ainsi, les nouveaux utilisateurs n'auraient pas la possibilité d'avoir des recommandations. Dans \cite{tagrecomm6}, Hotho \emph{et al.} ont proposé des recommandations de tags dans les \emph{folksonomies} basées sur les tags les plus utilisés. Cependant, ces recommandations ne sont absolument pas personnalisées étant donné que les mêmes tags sont proposés à chaque utilisateur justifiant par ailleurs l'introduction d'une dimension supplémentaire pour les \emph{folksonomies} dans notre approche, \emph{i.e.,} le profil des utilisateurs. Lipczak a proposé dans \cite{individusers} un système de recommandation de tags en trois étapes. \`{A} partir des tags annotés aux ressources, l'auteur ajoute des tags proposés par un lexique basé sur les co-occurrences de tags sur les mêmes ressources. Ensuite, le système filtre les tags déjà utilisés par l'utilisateur. Toutefois, malgré cette étape de filtrage, la recommandation ne paraît pas être personnalisée étant donné qu'elle cherche des tags co-occurrant sur d'autres annotations. L'approche revient ensuite à enlever les tags précédemment annotés par l'utilisateur de ceux qui sont suggérés. Dans \cite{landia}, les auteurs ont proposé une nouvelle approche combinant la similarité à la fois entre ressources et entre utilisateurs afin de recommander des tags personnalisés. En identifiant un ensemble d'utilisateurs similaire à l'utilisateur ciblé (en calculant leur similarité), l'approche est capable de suggérer des tags personnalisés pour l'utilisateur en question. En effet, deux utilisateurs sont considérés comme similaires s'ils ont assigné les mêmes tags aux mêmes ressources. Toutefois, il est rare de trouver pareille situation dans des \emph{folksonomies} où les tags utilisés par deux utilisateurs sur les mêmes ressources sont rarement identiques. Pour résumer, l'information (tags et ressources) considérée par ces travaux paraît incomplète afin de personnaliser au mieux les recommandations. Il nous paraît nécessaire de recourir à des informations supplémentaires sur le profil de l'utilisateur (profession, âge, etc.) afin de personnaliser au mieux les recommandations qui lui sont suggérés.

Notre objectif est donc d'utiliser des informations supplémentaires sur les utilisateurs afin de leur offrir une meilleure recommandation qui correspond à leur profil dans les \emph{folksonomies}. Nous allons montrer qu'en combinant ces données avec l'historique de tagging des utilisateurs, nous pouvons améliorer les recommandations. Par ailleurs, le problème majeur de la plupart des approches décrites est que les tags suggérés sont souvent ceux qui sont les plus utilisés dans les folksonomies. Ces tags ne sont cependant pas très spécifiques\footnote{Le problème est le même pour les ressources.}. Grâce aux concepts quadratiques, nous pouvons résoudre ce problème en mettant la lumière sur les tags qui, en plus d'avoir été partagés en masse, ont été utilisés en combinaison; le résultat est donc plus spécifique. Dans la section suivante, nous proposons une définition des concepts quadratiques qui permettent de regrouper l'information sur l'historique de tagging et sur le profil des utilisateurs. Par suite, nous introduisons notre système personnalisé de recommandations pour la suggestion de tags, de ressources et d'utilisateurs.

\section{Quadri-Concepts} \label{sec4}

Dans cette section, nous introduisons la notion de concept quadratique. Nous commençons par présenter une extension de la notion de \emph{folksonomie} \cite{Jaschke08} par l'ajout d'une quatrième dimension, \emph{i.e.,} le profil des utilisateurs.

\begin{definition}\label{dfolk} (\textsc{P-Folksonomie})
Une p-folksonomie est un ensemble de tuples $\mathcal{F}_{p}$ $=$ \textsc{(}$\mathcal{U}$, $\mathcal{T}$, $\mathcal{R}$, $\mathcal{P}$, $\mathrm{Y}$\textsc{)} où $\mathcal{U}$, $\mathcal{T}$, $\mathcal{R}$ et $\mathcal{P}$ sont des ensembles finis dont les éléments sont appelés \textbf{utilisateurs}, \textbf{tags}, \textbf{ressources} et \textbf{profils}. $\mathrm{Y}$ $\subseteq$ $\mathcal{U}$ $\times$ $\mathcal{T}$ $\times$ $\mathcal{R}$ $\times$ $\mathcal{P}$ représente une relation quadratique où chaque élément $y$ $\subseteq$ $\mathrm{Y}$ peut être représenté par un quadruplet : $y$ = \{($u$, $t$, $r$, $p$) $\mid$ $u$ $\in$ $\mathcal{U}$, $t$ $\in$ $\mathcal{T}$, $r$ $\in$ $\mathcal{R}$, $p$ $\in$ $\mathcal{P}$\} ce qui veut dire que l'utilisateur $u$ avec le profil $p$ a annoté la ressource $r$ via le tag $t$.
\end{definition}

\begin{remarque}
La dimension "profil de l'utilisateur" contient des données démographiques sur chacun des utilisateurs d'une p-folksonomie. Ces données peuvent concerner aussi bien l'âge que le genre ou encore la profession des utilisateurs. Plus tard, lors du processus de recommandation, nous nous intéresserons à l'\textbf{aspect évolutif} du profil, à savoir, les tags et ressources que partagent les utilisateurs.
\end{remarque}

Nous définissons maintenant un concept quadratique.

\begin{definition}\label{qc} (\textsc{Concept quadratique (fréquent)})
Un concept quadratique (ou quadri-concept) d'une p-folksonomie $\mathcal{F}_{p}$ = \textsc{(}$\mathcal{U}$, $\mathcal{T}$, $\mathcal{R}$, $\mathcal{P}$, $\mathrm{Y}$\textsc{)} est un quadruplet ($U$, $T$, $R$, $P$) avec $U$ $\subseteq $ $\mathcal{U}$, $T$ $\subseteq$ $\mathcal{T}$, $R$ $\subseteq$ $\mathcal{R}$ et $P$ $\subseteq$ $\mathcal{P}$ avec $U$ $\times$ $T$ $\times$ $R$ $\times$ $P$ $\subseteq$ $\mathrm{Y}$ tel que le quadruplet ($U$, $T$, $R$, $P$) est maximal, \emph{i.e.,} aucun de ces ensembles ne peut être augmenté sans diminuer un des trois autres ensembles. 
\end{definition}

Pour un \emph{quadri-concept} $qc$ = ($U$, $T$, $R$, $P$), les ensembles $U$, $R$, $T$ et $P$  sont respectivement appelés \textbf{\textit{Extent}}, \textbf{\textit{Intent}}, \textbf{\emph{Modus}} et \textbf{\textit{Variable}}.


\begin{remarque}
Sans restrictions de support, nous risquons d'avoir un grand nombre de concepts quadratiques pour une p-folksonomie donnée. Afin d'en garder les concepts les plus intéressants à étudier, nous devons définir des seuils minimaux de support sur chaque dimension de la p-folksonomie, i.e., \textit{$minsupp_u$}, \textit{$minsupp_t$}, \textit{$minsupp_r$} et \textit{$minsupp_p$}. Il en résulte des quadri-concepts qu'on appellera \textbf{fréquents}.
\end{remarque}

Afin de permettre l'extraction de l'ensemble de quadri-concepts fréquents à partir d'une \emph{p-folksonomie} donnée, nous pouvons utiliser l'un des deux algorithmes de la littérature dédiés à cette tâche: \textsc{QuadriCons} \cite{corr} ou \textsc{DataPeeler} \cite{datapeelertkdd}. Les deux algorithmes prennent en entrée une \emph{p-folksonomie} ainsi que quatre seuils minimaux de support sur chaque dimension et donnent en sortie l'ensemble de quadri-concepts fréquents vérifiant ces seuils.


Dans ce qui suit, nous utilisons les quadri-concepts afin d'évaluer notre approche sur un jeu de données du monde réel dans le but de voir comment les profils des utilisateurs peuvent être utilisés pour proposer une suggestion personnalisée de tags et de ressources.

\section{\textsc{PersoRec}: Un système personnalisé de recommandation pour les folksonomies}\label{sec5}

Dans cette section, nous introduisons notre algorithme de recommandation personnalisée pour les \emph{folksonomies}. Le pseudo-code de l'algorithme \textsc{PersoRec} est donné par l'Algorithme \ref{persorec}. \textsc{PersoRec} prend en entrée l'ensemble des quadri-concepts fréquents $\mathcal{QC}$ générés à partir d'une \emph{p-folksonomie}, un utilisateur $u$ avec son profil $p$ et (optionnellement) une ressource $r$ (à annoter) et donne en sortie trois différents ensembles: un ensemble d'utilisateurs proposés, un ensemble de tags suggérés et un ensemble de ressources recommandées.

\begin{algorithm}[htbp]\label{monalgo}
{

\Donnees{\begin{enumerate}  \item $\mathcal{QC}$: l'ensemble des quadri-concepts fréquents.
\item un utilisateur $u$ avec son profil $p$.
        \item  une ressource $r$.
\end{enumerate}}

\Res{ \begin{enumerate}   \item $\mathcal{PU}$: l'ensemble d'utilisateurs proposés.
        \item  $\mathcal{ST}$: l'ensemble des tags suggérés.
        \item  $\mathcal{RR}$: l'ensemble des ressources recommandées.
\end{enumerate}}

    \Deb {

\PourCh {quadri-concept $qc$ $\in$ $\mathcal{QC}$ } {
\Si{$p$ $\in$ $qc.Variable$}{
/*\emph{Proposition d'utilisateurs}*/\\
\PourCh {utilisateur de $qc.extent$}
{
\Si{$qc.extent$ $\neq$ $u$}{
$\mathcal{PU}$ = $\mathcal{PU}$ $\cup$ $qc.extent$;
}
}

/*\emph{Suggestion de tags}*/\\
\Si{$qc.Intent$ = $r$}{
$\mathcal{ST}$ = $\mathcal{ST}$ $\cup$ $qc.modus$;
}

/*\emph{Recommandation de ressources}*/\\
$\mathcal{RR}$ = $\mathcal{RR}$ $\cup$ $qc.Intent$;
}

}
}
\Retour ($\mathcal{PU}$,$\mathcal{ST}$,$\mathcal{RR}$); }

  \caption{\textsc{PersoRec}}
  \label{persorec}
\end{algorithm}

\textsc{PersoRec} parcourt l'ensemble des quadri-concepts fréquents afin de chercher ceux où le profil $p$ de l'utilisateur $u$ appartient à la \emph{variable} (\emph{i.e.,} l'ensemble de profils). Ensuite, selon le type d'application souhaitée, \textsc{PersoRec} opère comme suit:

\begin{description}
  \item[Proposition d'utilisateurs] Pour être en mesure de recommander à un utilisateur donné une liste personnalisée d'utilisateurs ayant un profil et des intérêts équivalents, nous exploitons les quadri-concepts et plus particulièrement l'information contenue dans la quatrième dimension. Ainsi, \textsc{PersoRec} cherche les quadri-concepts dont les utilisateurs ont le même profil que $u$. Si $u$ appartient déjà à un quadri-concept trouvé, il est filtré, sinon, il est ajouté à l'ensemble $\mathcal{PU}$ des utilisateurs proposés (Ligne $7$).
  \item[Suggestion de tags]  L'objectif est de suggérer à un utilisateur, qui désire partager une ressource dans la \emph{p-folksonomie}, une liste personnalisée de tags. Pour ce faire, nous lui suggérons les tags qui ont été affectés à cette même ressource par des utilisateurs au profils équivalents. Le but étant de recommander à $u$ des tags qui sont le plus en concordance avec son profil. Ainsi, pour cette application, nous avons besoin d'une information supplémentaire, \emph{i.e.,} la ressource à annoter ($r$). Par suite, nous ajoutons les tags affectés à cette ressource $r$ par des utilisateurs au même profil que $u$ à l'ensemble $\mathcal{ST}$ (Ligne $10$).
  \item[Recommandation de ressources] Le but de cette perspective est de recommander une liste personnalisée de ressources, à un utilisateur $u$, susceptible de correspondre à ses intérêts. Pour ce faire, à partir du profil de $u$, nous cherchons les ressources à partir des quadri-concepts correspondant à ce profil. Ainsi, \textsc{PersoRec} est capable de recommander des resources qui ont été partagées par des utilisateurs ayant le même profil que $u$. L'ensemble $\mathcal{RR}$ contient donc ces ressources qui seront recommandées à $u$ (Ligne $12$).
\end{description}


Dans ce qui suit, nous évaluons notre approche sur un jeu de données du monde réel: \textsc{MovieLens}. Nous allons donner quelques exemples intéressants de quadri-concepts et de recommandations puis nous évaluerons la qualité de ces recommandations, \emph{i.e.,} la précision de ces dernières.

\section{\'{E}tude expérimentale} \label{sec6}

Dans cette section, nous évaluons notre approche sur un jeu de données du monde réel: \textsc{MovieLens} en donnant quelque exemples de quadri-concepts extraits selon les différents profils des utilisateurs. \textsc{MovieLens} est un système de recommandation qui permet aux utilisateurs d'annoter des films. Le jeu de données utilisé \cite{movie}\footnote{http://www.grouplens.org/node/73} contient $95580$ tags affectés à $10681$ films par $71567$ utilisateurs. Des informations supplémentaires sur les utilisateurs sont disponibles dans le jeu de données et forment son profil (la quatrième dimension d'une \emph{p-folksonomie}) et qui renseigne sur le \textbf{genre} de l'utilisateur (masculin ou féminin), sa \textbf{profession} (au nombre de $21$, qui peut être éducateur, écrivain, étudiant, scientifique, etc.) ainsi que sur son \textbf{âge} ($5$ tranches d'âge).




\subsection{Exemples de quadri-concepts extraits}



Dans ce qui suit, nous présentons quelques résultats intéressants de quadri-concepts extraits par \textsc{Quadricons} à partir du jeu de données \textsc{MovieLens}. Pour une telle extraction, nous avons exécuté les deux algorithmes de la littérature dédiés à cette tâche, \emph{i.e.,} \textsc{QuadriCons} et \textsc{Data Peeler} sur une machine munie d'un processeur Intel Core $i7$ avec une mémoire de $4$ Go. Différents tests, réalisés sur le système d'exploitation Linux Ubuntu $10$.$10$.$1$ ($64$ bits), ont démontré que l'algorithme \textsc{QuadriCons} est plus performant que son concurrent en temps d'exécution (\emph{e.g.,} entre $1$ et $50$ secondes pour \textsc{QuadriCons} contre $300$ secondes au minimum pour \textsc{Data Peeler}). \'{E}tant donné la grande taille du jeu de données considéré, nous avons donc opté pour l'algorithme \textsc{QuadriCons} afin de générer les quadri-concepts fréquents <utilisateurs, tags, ressources, profils>.

Nous avons défini les valeurs de seuils de supports suivants: \textit{$minsupp_u$} = $2$, \textit{$minsupp_t$} = $2$, \textit{$minsupp_r$} = $2$ et \textit{$minsupp_p$} = $2$\footnote{Dans un quadri-concept fréquent, $2$ utilisateurs (au moins) avec deux mêmes profils (\emph{e.g.,} même profession et même age) ont assigné les mêmes tags ($2$ au moins) aux mêmes ressources ($2$ au moins).}. De toute évidence, il est plus intéressant de fixer chaque seuil de support à $2$ dans le but d'avoir des quadri-concepts avec une valeur ajoutée illustrant les tags et ressources partagés en commun par un groupe de deux utilisateurs (au moins) ayant le même profil. Ainsi, le Tableau \ref{all} illustre quelques exemples (des plus intéressants) de quadri-concepts parmi les $10627$ quadri-concepts fréquents vérifiant les seuils de supports décrits ci-dessus. Par exemple, e premier quadri-concept montre que les utilisateurs \emph{bernadette}, \emph{bridget} et \emph{margaret62}, trois femmes à la retraite, ont partagé les films \emph{Star Wars}, \emph{M.A.S.H} et \emph{Rear Window} via les tags \emph{classic}, \emph{dialog} et \emph{oscar}. Dans le deuxième quadri-concept, nous pouvons observer que les hommes travaillant dans le domaine de la santé ont assigné les tags \emph{cult} et \emph{bestmovie} aux trois films décrits par le Tableau \ref{all}.

\begin{table*}[htbp]
\begin{center}
{

\begin{tabular}{|r|r|r|r|}
\hline
     Utilisateurs &       Tags &  Ressources &    Profil \\
\hline

bernadette &    classic &  Star Wars &            \\
\hline
   bridget &     dialog &    M.A.S.H & Femme, 46-73 ans, retraité \\
\hline
   margaret62 &      oscar & Rear Window &            \\
\hline
\hline
    mulder &        bestmovie    & Usual Suspects &            \\
\hline
    scully &       cult & Silence of the Lambs & 25-35 ans, Homme, domaine santé \\
\hline
  csmdavis &   & Sound of Music &            \\
  \hline
\hline
     rossy &    classic & Rear Window &          \\
\hline
   anlucia &   oldmovie & Magician of OZ &   36-45 ans, Homme, écrivain    \\
   \hline
   franela &   quotes & Gone with the Wind &     \\
  \hline
\hline

\end{tabular}}

\end{center}
\caption{Exemples de quadri-concepts extraits à partir du jeu de données \textsc{MovieLens}.\label{all}}
\end{table*}

\subsection{Recommandation personnalisée}

Les quadri-concepts on permis de regrouper sous un même concept des utilisateurs au profil équivalent qui partagent des tags et ressources en commun. Les quadri-concepts peuvent être appliqués dans plusieurs domaines dont la suggestion de tags, la recommandation de ressources ou la proposition d'amis grâce à l'algorithme \textsc{PersoRec}.\\

\textbf{Suggestion de tags:} Par exemple, considérons l'utilisateur \emph{rossy} qui souhaite partager le film \emph{Rear Window}: puisque \emph{rossy} est âgé entre $36$ et $45$ ans, il aurait la possibilité d'utiliser les tags \emph{classic}, \emph{oldmovie} et \emph{quotes}, par contre, s'il était âgé, par exemple, de $60$ ans, nous lui aurions proposé les tags \emph{classic}, \emph{dialog} et \emph{oscar}. De même, selon la profession de \emph{davis}, les tags suggérés seront différents pour correspondre au mieux à son vocabulaire.

\textbf{Recommandation de ressources:} Cela semble trivial de dire qu'un jeune utilisateur sera plus intéressé par les comédies que par des films classiques; de même, il serait plus judicieux de recommander des films romantiques à une femme qu'un film de guerre. Dans notre cas, considérons deux nouveaux utilisateurs de la \emph{folksonmie} \emph{reyes} ($51$ ans) et \emph{zlatan} ($26$ ans). Contrairement aux systèmes de recommandation classiques, il sera possible grâce à notre système de recommander des films à ces nouveaux utilisateurs en dépit de l'absence d'historique de tagging. Ainsi, nous proposerons les films \emph{Star Wars}, \emph{M.A.S.H} et \emph{Rear Window} à \emph{reyes} tandis que les films \emph{Usual Suspects}, \emph{Silence of the Lambs} et \emph{Sound of Music}, plus susceptibles de plaire à \emph{zlatan}, seront recommandés à ce dernier. Le choix inverse aurait été beaucoup moins pertinent.

\textbf{Proposition d'amis:} Par exemple, si nous considérons l'utilisateur \emph{csmdavis} qui est médecin (\emph{cf.,} Tableau \ref{all}), nous pourrons lui recommander comme amis les utilisateurs \emph{mulder} et \emph{scully} puisqu'ils ont des profils et intérêts équivalents, \emph{i.e.,} tous les trois médecins et appartenant à la même tranche d'âge, en plus d'être intéressés par les mêmes tags et ressources.

\begin{figure}[htbp]
\begin{center}
\includegraphics[scale=0.5]{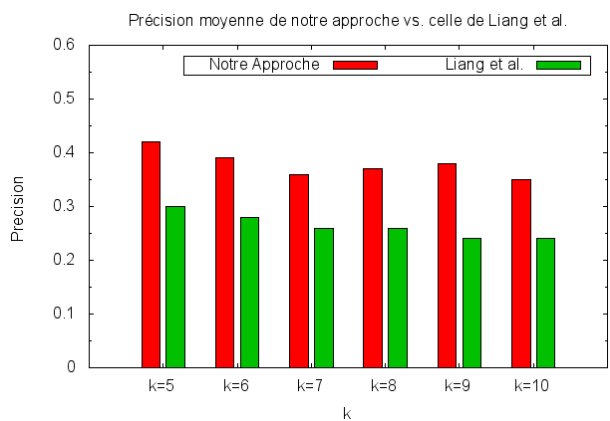}
\end{center}
\vspace{-0.4cm} \caption{Précision moyenne de notre approche pour la recommandation de ressources \emph{vs.} celle de Liang et al.}\label{precision2}
\end{figure}

\subsection{\'{E}valuation de l'approche}

\'{E}valuer l'efficacité d'un algorithme de recommandation est loin d'être trivial. En premier lieu, parce que différents algorithmes peuvent être meilleurs ou moins bons en fonction du jeu de données sur lequel ils sont appliqués. D'autre part, les objectifs fixés par un système de recommandation peuvent être divers et variés. Un système de recommandation peut être mis en place pour estimer avec exactitude la note que donnerait un utilisateur à un élément, alors que d'autres auront comme objectif principal de ne pas proposer des recommandations erronées. On peut donc légitimement se demander jusqu'à quel point ces différentes méthodes de recommandation sont réellement efficaces. Néanmoins, pour déterminer l'efficacité d'un système, l'indicateur le plus répandu dans la littérature est la précision (\emph{cf.,} Equation \ref{eq}) qui représente la qualité de la recommandation, c'est-à-dire à quel point les suggestions proposées sont conformes aux intérêts de l'utilisateur.
\begin{equation}\label{eq}
\emph{Précision} = \frac{nombre~de~recommandations~pertinentes}{nombre~de~recommandations}
\end{equation}
La précision détermine donc la probabilité qu'un élément recommandé soit pertinent. Ainsi, la meilleure mesure de l'efficacité d'un algorithme de recommandation et de la pertinence des suggestions est donc d'évaluer la précision de la prédiction effectuée par le système en comparant les prédictions avec les choix qu'aurait fourni l'utilisateur dans le cas réel \cite{apprentissage}. Dans nos expérimentations, nous avons également fait varier le nombre de recommandations proposées à l'utilisateur: on parle alors de requête top-$k$. Grâce à ce genre de requête, l'utilisateur peut spécifier le nombre $k$ de réponses (recommandations) les plus pertinentes que le système doit lui retourner. Cela permet surtout d'éviter de submerger l'utilisateur par un grand nombre de réponses en lui retournant que le nombre de réponses les plus pertinentes qu'il souhaite \cite{topk}. Dans ce qui suit, nous nous intéressons à une des trois applications de notre système de recommandation, \emph{i.e.,} la recommandation de ressources et nous évaluons la précision de notre approche \emph{vs.} l'approche correspondant le plus à nos travaux, \emph{i.e.,} celle de Liang \emph{et al.} \cite{Liangarticle} qui génére des recommandations personnalisées basées sur des informations sur les utilisateurs. Ainsi, la Figure \ref{precision2} montre les différentes valeurs de précision obtenues par notre algorithme de recommandation \emph{vs.} celui de Liang \emph{et al.} pour différents valeurs de $k$ allant de $5$ à $10$. De manière générale, les recommandations faites aux utilisateurs de \textsc{MovieLens} correspond aux attentes de ces derniers, \emph{i.e.,} les recommandations sont pertinentes à un niveau moyen de $38\%$ ce qui correspond à un score meilleur que celle de l'approche de Liang \emph{et al.} which is between $24\%$ and $30\%$. Cela suggère que le recours aux quadri-concepts améliore les recommandations en suggérant les tags et ressources les plus adéquats aux besoins des utilisateurs. Les résultats obtenus montrent, par ailleurs, que les meilleures performances sont obtenues avec une valeur de $k$ = $5$. Cela est dû au fait que les $5$ premières recommandations correspondent aux attentes des utilisateurs et lorsque le nombre de recommandations augmente, cela entraîne inévitablement une diminution de la précision étant donné que l'utilisateur choisit moins de ressources que celles qui lui sont recommandées.

\section{Conclusion et perspectives} \label{sec7}

En raison de sa simplicité d'utilisation, les \emph{folksonomies} ont très rapidement attiré des millions d'utilisateurs qui partagent des ressources en leur affectant des tags. Dans ce papier, nous avons considéré donc le profil des utilisateurs comme une nouvelle dimension d'une \emph{folksonomie} et nous avons utilisé l'algorithme \textsc{QuadriCons} afin de permettre l'extraction des quadri-concepts formés d'utilisateurs, de tags, de ressources et de profils. Ces concepts seront ensuite utilisés afin d'offrir un choix de tags et de ressources plus personnalisé correspondant précisément au profil des utilisateurs. Parmi les perspectives pour nos travaux futurs, nous pouvons enquêter sur le \emph{feedback} des utilisateurs afin de montrer si nos suggestions sont mieux considérés que les suggestions brutes. Nous pourrons également explorer d'autres cadres applicatifs qui utilisent les quadri-concepts comme l'auto-complétion de requêtes \cite{coria10}. Enfin, il sera intéressant d'étudier le récent domaine de la recherche sociale.

\small{
\bibliographystyle{abbrv}

\bibliography{biblio}
}

\end{document}